\documentstyle[12pt,myown]{article}
\newcommand{\beq}{\begin{equation}}
\newcommand{\eeq}{\end{equation}}
\title{\large \sf \begin{flushright} Report NPI Rez -- EXP -- 01/98
    \end{flushright}
\vspace{5mm}
 \bf A possible origin of the endpoint anomaly in tritium 
$\beta$-spectrum}
\author{\normalsize J.\v R\'{\i}zek, V.Brabec, O.Dragoun, 
M.Ry\v sav\'y\thanks{e-mail: rysavy@ujf.cas.cz},
    A.\v Spalek \\
\small \it Nuclear Physics Institute, Acad. Sci. of Czech Republic,
    \vspace{-0.7ex}\\
\small \it CZ-250 68 \v Re\v z near Prague, Czech Republic}
\date{ }

\begin{document}

\maketitle
\begin{abstract}
The influence of the residual T atoms appearing after the decay of
T$_2$ molecule on the $\beta$-spectrum shape is considered. Recent
experiments performed in Mainz, Troitsk, and Livermore are briefly
reviewed from this viewpoint. Aspects connected
with the possible time dependent change of the tritium source
composition are discussed.
\end{abstract}

\section{Introduction}
The recent attempts to determine the limits
on the electron antineutrino mass from the  endpoint
part of the $\beta$-spectrum of  tritium decay revealed a strange fact.
When fitting the measured spectrum in all the  experiments
\cite{Rob91} -- \cite{Bele95} %[1 - 6]
performed in this decade,
the negative (i.e. nonphysical) values of the antineutrino-mass squared
were obtained. The reason for this is presently unknown.
The recent review of the negative neutrino-mass  squared problem
and some considerations are given in \cite{Kap97}. %ref.[7].

In this note  some remarks concerning the measurement
and interpretation of the tritium  $\beta$-spectrum
are given. We consider the case when the radioactive source is composed
of the $T_{2},\;(R_{0},T)\;{\rm and} \;(R_{1},R_{2})$ molecules.
Here $R_{i}$ denote the nonradioactive parts of the molecules.
We take into account that after the decay of one $T$
atom in the $T_{2}$ molecule  the other atom  can remain  in the
source and its decay can influence the shape of the
measured $\beta$-spectrum.

Let us at first consider the experiments carried by
the Mainz \cite{Wein93,Bac96} %[5,8]
 and Troitsk \cite{Bele95,Lob96} %[6,9]
 groups, and the result of Livermore (LLNL) experiment \cite{Sto95}. %[10].

\section{Results of Troitsk, Mainz and LLNL experiments}
The Troitsk experiment consists at present  from two runs - 1994 and 1996.
In the 1994 run the negative $m_{\nu}^{2}$
was obtained \cite{Bele95}. As a reason for this nonphysical result,
 the excess of counts at
$7.6\;{\rm eV}$ below the endpoint was claimed. When corrected
for this excess, which was interpreted as a monoenergetical bump in
the differential spectrum with branching ratio of about
$7 \times 10^{-11}$,
the negative value of $m_{\nu}^{2}$ diminished.
Thus in this 1994 experiment, instead of the puzzle of the negative
$m_{\nu}^{2}$, another one -- a strange anomaly in the $\beta$-spectrum
near the endpoint -- appeared. The value of $m_{\nu}^{2}$ was
found to depend  considerably on the size of the fitted
energy interval below the endpoint.
When the interval was sufficiently large, the
negative antineutrino mass squared appeared again. This was explained
by the excess of counts below $18300\;eV$ energy. The energies and
populations of molecular final states were taken into account
using the results of theoretical calculations \cite{Fac85,Kol88}. %[11,12].

The 1996 experiment \cite{Lob96} confirmed the main conclusions of
the previous one and led to new problems.
The interpretation of the data was improved in the sense that the
effect of the trapping the electrons in the
tritium source was treated more carefully
(it turned out that this effect was underestimated earlier).
Another set of final states was  taken into
the account in accordance with the newest theoretical
results \cite{Jon96}. %[13].
However, the position of the bump-like anomaly turned out to be time
dependent: in the 1996 run it was situated at $12.3\;eV$
below the endpoint while its intensity remained almost
the same as in previous run. Taking into account all the
corrections and the bump-like structure, the fitted value of the
$m_{\nu}^{2}$ was positive. But again the result depended on the
size of the energy interval considered. In  Table 1 the results of
Troitsk experiment are summarized.

\vspace{5mm}

\begin{table}[h]
\caption{ Summary of the Troitsk experiment}
\vspace{3mm}
\begin{tabular}{|c|c|c|c|c|c|}
\hline
$m_{\nu}^{2}\;(eV)^{2}$&$m_{\nu}\;(eV)$&
endpoint ({\it eV})&interval from ({\it eV})&run&{\tiny 2)}\\
\hline
 $-22\pm4.8$ & & & 18350 & 1994 &no\\
\hline
 $-4.1\pm10.9$ & & 18575 & & 1994&yes\\
\hline
$-2.7\pm10.1\pm4.9\;^{1)}$&$<4.35$&
\ \ 18575.4&18300&1994&yes\\
\hline
$-11.\pm5.4$& & &18300&1996&yes\\
\hline
$3.8\pm7.4\pm2.85$&$<4.4$&18575&
18400&1996&yes\\
\hline
$1.5\pm5.9\pm3.6$&$<3.9$&18575&
&1994+1996 &yes\\
\hline
\end{tabular}
%\vspace{2mm}
$^{1)}$ corrected in 1996
\newline
$^{2)}$ indicates whether or not the spectrum was corrected for
the near endpoint anomaly
\end{table}

\vspace{5mm}

As concerned the Mainz experiment, there were also two runs -- in 1991
and in 1994. No  endpoint anomaly was reported. However, as in the Troitsk
experiment, an excess of counts at the energies
of $\sim$100$\;eV$ below the endpoint
was observed. The results of Mainz experiment are summarized in Table 2.

\vspace{5mm}
\begin{table}[h]
\caption{Summary of Mainz experiment}
\begin{center}
\begin{tabular}{|c|c|c|c|c|}
\hline
$m_{\nu}^{2}\;(eV)^{2}$&$m_{\nu}\;(eV)$&endpoint $(eV)$&
size of interval $(eV)$&run\\
\hline
$-39\pm34\pm15$&$<7.2$&18574.8&140 &1991\\
\hline
$-22\pm17\pm14$&$<5.6$& & 143 &1994\\
\hline
$-5\pm27\pm5$&$<7.1$& &100 &1994\\
\hline
\end{tabular}
\end{center}
\end{table}
\vspace{5mm}

Different tritium sources were used in the above
experiments -- gaseous one in Troitsk and condensed molecular film in Mainz.
According to  ref.\cite{Bele95}, %[6]
 the Troitsk source consisted in 1994 from the mixture of
$T_{2}+\left(HT\right)+H_{2}$ in the proportion of 6:8:2. Some
admixture of free $T^{-}$ ions was also reported but considered to be
negligible. The composition of the Mainz source was, in accordance
with ref.\cite{Przy95}, %[14],
73.9\% $T_{2}$, 19.2\% $(H,T)$ and 6.9\% $H_{2}$.

Finally let us cite the results of ref.\cite{Sto95}. %[10].
In that LLNL experiment, the gaseous source of molecular tritium was used,
the purity of which was declared to be 98\%. Anomalous structure in the
last 55 {\it eV} below the endpoint was  reported. This was interpreted
as a narrow peak at $\sim$23$\;eV$ below the
endpoint with intensity $3 \times 10^{-9}$ of the total $\beta$-strength.

The question arises whether the mentioned anomalies are due to the
fact that  especially the tritium decay is studied or whether
it can be observed in other decays,
too. Unfortunately the problem of the final atomic or molecular states
makes it  very difficult to study this problem
using more complicated atoms or molecules than the tritium ones.
In addition, the interpretation has to rely only upon the theoretical
calculations of the final states energies and populations.
No experimental data on the  final states are available.
Many reasons that can cause these observed anomalies were qualitatively
discussed -- {\it e.g.} account of some missed final states, possible 
neutrino mixing, special neutrino interaction.

Especially intriguing is the observed time dependence of the near endpoint
count excess. It seems that the composition
of the source can  change with time.
In the above mentioned three experiments the radioactive sources of different
compositions were used.
A decay of one tritium atom of the T$_2$ molecule changes the composition
of the source and this change \underline{\it is} time dependent.
In what follows we try to analyse qualitatively namely this point.

\section{The decay of $T_{2}$}
The decay of molecular tritium
proceeds as follows:
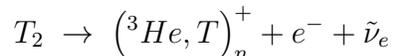
\[T_{2}\;\rightarrow\;
\left(^{3}He,T\right)^{+}_{n}
+e^{-}+\tilde{\nu}_{e}\]
Here $n$ denotes the different molecular branches (final states) of the decay.

Probably in most cases the $\left(^{3}He,T\right)^{+}$ ion breaks
and the tritium atom  reacts with some residual atom or with the
atoms of wall surrounding the initial $T_{2}$ molecules.
It seems possible that some number of new molecules with the tritium atom
as a part will remain in the radioactive source and will contribute to
measured spectrum. In the next we consider the decay more quantitatively.

Let us suppose that at time $t=0$ we have 100\% of $T_{2}$
molecules in the source. Further on we will use the following
notation:
\begin{description}
\item[ ] $N_{2}\left(t\right)\;\cdots
\;$ {\rm the number of} $T_{2}$
{\rm molecules at time} $t$,
\item[ ] $N_{1}\left(t\right)\;\cdots\;$
{\rm the number of the tritium atoms
that are not a part of the $T_{2}$ molecule},
\item[ ] $N\left(t\right)\;\cdots\;$
{\rm the whole number of the tritium atoms
at time} $t$.
\end{description}
It is clear that the following relations
hold ($\lambda$ denotes the tritium decay constant):
\beq
N\left(t\right)
=2N_{2}\left(0
\right){\rm e}^{-\lambda t}=N_{1}
\left(t\right)+2N_{2}\left(t
\right)
\eeq
\beq
N_{1}\left(t+\Delta t\right)
-N_{1}\left(t\right)=
-\lambda N_{1}\left(t\right)\Delta t
+\left(N_{2}\left(t\right)-
N_{2}\left(t+\Delta t\right)\right)
\eeq

From (1) we can get the relation between derivatives
of $N_{1}\left(t\right)\;{\rm and}\; N_{2}\left(t\right)$, from the equation
(2) we obtain (in the limit $\Delta t\rightarrow\;0$)
another relation between these derivatives. As a result the differential
equations for $N_{1}\;\left(t\right){\rm and}\;N_{2}
\;\left(t\right)$ are obtained. Explicitly this is written as

\begin{eqnarray}
-\frac{dN_{2}\left(t\right)}{dt}&=&
\frac{1}{2}\left[\frac{dN_{1}\left(t\right)}{dt}
+2\lambda N_{2}\left(0\right)
e^{-\lambda t}\right],\nonumber\\
\frac{dN_{1}\left(t\right)}{dt}&=&
-\lambda N_{1}\left(t\right)-
\frac{dN_{2}\left(t\right)}{dt}.
\end{eqnarray}
Using these relations we get the equation for $N_{1}\left(t\right)$
\beq
\frac{dN_{1}\left(t\right)}{dt}
=-2\lambda\left[ N_{1}\left(t\right)
-N_{2}\left(0\right){\rm e}^{-\lambda t}\right],
\;\;N_{1}\left(0\right)=0.
\eeq
The solution fulfilling initial condition is
\beq
N_{1}\left(t\right)=2N_{2}\left(0\right)
\left[1-{\rm e}^{-\lambda t}\right]{\rm e}^{-\lambda t},
\;\;N_{2}\left(t\right)=N_{2}\left(0\right)
{\rm e}^{-2\lambda t}.
\eeq
Therefore we have
\beq
\frac{N_{1}\left(t\right)}
{N_{2}\left(t\right)}=2
\frac{1-{\rm e}^{-\lambda t}}{{\rm e}^{-\lambda t}}.
\eeq
This is the fraction of the tritium atoms with respect to
the number of $T_{2}$ molecules that are not part of the tritium
molecules. For illustration, this fraction amounts
to 0.3\% and 10\% for t equal 10 days and one year, respectively.
The  question is where these atoms (which undoubtedly
appear) are in a real experiment.
Are they removed in some way in order not to influence the measurement
or they remained within the source in a form different from
the molecular tritium? They can be as well  absorbed
on the walls of the bottle where the original molecular tritium is
stored. Of course the fraction will be different from that
predicted by eq.(6) if some  of the tritium atoms escape out of the source.

In the more general case when the initial number of tritium
atoms is not zero we have
\beq
N_{1}\left(t\right)=
N_{1}\left(0\right){\rm e}^{-\lambda t}+
2N_{2}\left(0\right)\left(1-
{\rm e}^{-\lambda t}\right){\rm e}^{-\lambda t},
\;\;N_{2}\left(t\right)=N_{2}
\left(0\right){\rm e}^{-2\lambda t}.
\eeq
For completeness let us give the results for the source which composition is
\beq
N_{2}\left(0\right)=p_{2}N\left(0\right),
\;N_{1}\left(0\right)=p_{1}N\left(0\right),
\;N_{0}\left(0\right)=p_{0}N\left(0\right),
\eeq
where $N_{i}\left(0\right)$ denotes the number of molecules
T$_{2}${\it (i=2)}, RT {\it (i=1)} and R$_{1}$R$_{2}$ {\it (i=0)}
at time $t=0$. Then the following relations at time $t$ hold
\begin{eqnarray*}
N_{2}\left(t\right)&=&p_{2}N\left(0\right)
{\rm e}^{-2\lambda t},\\
N_{1}\left(t\right)&=&
N\left(0\right)\left[p_{1}+
2p_{2}\left(
1-{\rm e}^{-\lambda t}\right)
\right]{\rm e}^{-\lambda t},\\
N_{0}\left(t\right)&=&
N\left(0\right)\left[p_{0}
+p_{1}\left(
1-{\rm e}^{-\lambda t}\right)+
p_{2}\left(
1-{\rm e}^{-2\lambda t}\right)\right].
\end{eqnarray*}
In the two next sections we consider some aspects that can be connected
with different composition of the tritium source.

\section{Different endpoint energies}
From the above consideration we  can expect that some,
may be unknown, admixture of type $(T,R)$ will be present in the source.

Let us suppose that this is the case
and consider the resulting $\beta$-spectrum as a superposition of two spectra.
The essential here is that the energy endpoints for these two branches are
different. We consider the situation near endpoint  and suppose that
the neutrino mass is zero. The expression
$F\left(Z,E\right)E\sqrt{E^{2}-1}$ (here $E$ denotes the full electron
energy ($\hbar=m=c=1$), $F$ stands for the Fermi function)
turns out to be almost constant on the 300 $eV$ interval
below endpoint. In order to simplify the calculation we suppose that
it {\it is} a constant. Then the differential spectrum near the endpoint
energy is written as follows:
\beq
\frac{dP\left(E\right)}{dE}=
C\left\{\begin{array}{cc}
A\left[W_{0}-E\right]^{2}&W_{0}
\geq E\geq W_{0}-\Delta E,\\
A\left[W_{0}-E\right]^{2}+
B\left[W_{0}-E-\Delta E\right]^{2}&
E\leq W_{0}-\Delta E.
\end{array}\right.
\eeq
As is seen we consider the superposition of two spectra
with amplitudes $A,\;B$ ($A + B = 1$) and endpoint energies $W_{0},
\;W_{0}-\Delta E$, respectively.
The normalization constant is denoted as $C$. The integral spectrum from
some $E_{low}<W_{0}-\Delta E$ up to $W_{0}$ is
\beq
I\left(E_{low}\right)=C\left\{
\frac{1}{3}y^{3}-B\Delta E \Phi
\left(y\right)\right\}
\eeq
where
\[y=W_{0}-E_{low},\;\;
\Phi\left(y\right)=
y^{2}-\Delta E y+\frac{1}{3}\left(\Delta E
\right)^{2}.\]

Usually several measurements are done that are separated from each other
by some time interval. From our viewpoint
the constants $B$ {\it are different in these measurements due to the
different composition of the sources}. If we normalize the two measurements
to the same integral intensity at some point $y_{0}$, we get
for the normalization factors the relation
\beq
\frac{1}{3}\left(C_{1}-C_{2}\right)y_{0}^{3}=
\left(B_{1}C_{1}-B_{2}C_{2}\right)
\Delta E\;\Phi\left(y_{0}\right).
\eeq
Using this relation we can write for the difference of two the spectra
\beq
I_{1}\left(y\right)-I_{2}\left(y\right)=
\frac{1}{3}\left(C_{1}-C_{2}\right)
\left\{y^{3}-\frac{y_{0}^{3}}
{\Phi\left(y_{0}\right)}
\Phi\left(y\right)\right\}.
\eeq
This function has local extrema at the points
\beq
y_{\pm}=\frac{y_{0}^{3}}
{3\Phi\left(y_{0}\right)}\left(
1\pm\sqrt{1-3\Delta E\frac{
\Phi\left(y_{0}\right)}{
y_{0}^{3}}}\;\right).
\eeq
Only one point is feasible due to the condition $y > \Delta E$.
Of course the results obtained are qualitative only. We
neglect the energy dependence arising from the common factor including
the Fermi function (in accordance with our numerical calculations
this dependence is linear but close to a constant in the interval
considered).  We also know nothing about the actual ``fate'' of
the residuals after one of the tritium atom from the molecule decays.
But if these residuals will in part create some impurities
in the source there is a possibility that these can serve as a reason
of observed anomalies in the spectra near the endpoint.

If we consider the expression for the spectra
with $B\neq0$,  we see that under the preposition of small
both $B\;{\rm and}\; \Delta E$ the part of the spectrum influenced
by this term is very narrow and concentrated somewhere below
$W_{0}-\Delta E$. This term will be quickly embedded in the
statistical error of dominating
part connected with $A$ branch of the spectrum when going sufficiently
below this energy. Let us notice that the ``effective position'' where
this term is in effect depends on $B$ and therefore it is --
from our viewpoint -- time dependent. The question is whether this influence
can be simulated by some step function at some energy.

We can also point out  another aspect. Namely, let us consider
the case of nonzero neutrino mass and suppose that $B=0$.
Then the integral spectrum will be (we consider this spectrum from such
electron energies that the neutrino mass may be considered small
and the approximation
$\sqrt{1-\frac
{m_{\nu}^{2}}{\left(W_{0}-E\right)^{2}}}
\simeq 1-\frac{m_{\nu}^{2}}{
2\left(W_{0}-E\right)^{2}}$
can be used)
\beq
I\left(E\right)=C\left\{\frac{1}{3}y^{3}
-\frac{1}{2}m_{\nu}^{2}y+\frac{1}{2}
m_{\nu}^{2}\delta E\right\},\;\;y=W_{0}-E.
\eeq
Here $\delta E$ denotes such energy interval below $W_{0}$
 that we can use the approximation of the square root written above
(i.e. we integrate from $E$ to $W_{0}-\delta E$).
If we now compare this expression with the spectrum for the case
$B\neq0,\;m_{\nu}=0$, eq. (10), we see that these spectra have  common
main term but differ in
\begin{eqnarray}
-B\Delta E y^{2} +B\left(\Delta E\right)^{2}y
-\frac{1}{3}B\left(\Delta E\right)^{3}&{\rm in}&
\;(10)\nonumber\\
-\frac{1}{2}m_{\nu}^{2}y+\frac{1}{2}m_{\nu}
^{2}\delta E&{\rm in}&\;(14)
\end{eqnarray}
If the spectrum measured is indeed that of (10) and fitted one that of (14)
we see that we can expect to find the nonzero square of neutrino mass.
It is probable that the fitted squared neutrino mass will be positive
unless we have a term $\sim$$y^{2}$ to fit the background.

Let us notice that we can have the case when the endpoint
energy of the $B$ branch lies higher than that of $A$ branch.
The same considerations as given above result in the following conclusions:
\begin{enumerate}
\item The difference in spectra
leads to the local extremum
at the point
\beq
y_{+}=\frac{y_{0}^{3}}
{3\Psi\left(y_{0}\right)}
\left\{1+\sqrt{1+3\Delta E
\frac{\Psi\left(y_{0}\right)}
{y_{0}^{3}}}\;\right\}
\eeq
\noindent where
\[\Psi\left(y\right)=
y^{2}+\Delta E y+\frac{1}{3}
\left(\Delta E\right)^{2}.\]
\item The different term in analogy with the consideration
of zero and nonzero neutrino mass as given above in equation (15) is:
\[B\left(\Delta E y^{2}+\left(\Delta E\right)^{2}y
+\frac{1}{3}\left(\Delta E\right)^{3}\right).\]
\noindent We see that in this case we obtain, when fitting neutrino mass,
negative value for $m_{\nu}^{2}$.
\end{enumerate}
When the difference of the $\beta$-spectra taken in Troitsk in  1994
and 1996  were investigated some maximum was indeed found.
\section{The same endpoint}
The theoretical spectrum is  described
by the following formula (again we suppose that the above mentioned factor
is constant or can be eliminated as in the case when Kurie plot is
considered)
\beq
P\left(E\right)=\sum_{n}w_{n}
\left(E_{0}-E-\varepsilon_{n}\right)
\sqrt{\left(E_{0}-E-\varepsilon_{n}
\right)^{2}-m_{\nu}^{2}}\;\;
\theta\left(E_{0}-E-\varepsilon_{n}-m_{\nu}
\right).
\eeq
Here $E_{0},\;E,\; w_{n},\;\varepsilon_{n}$ denote the endpoint energy,
the electron energy, probability of the excitation of
$n$-th ion excited state and its energy, respectively.
For small $m_{\nu}$ we get
\begin{eqnarray}
P\left(E\right)&=&W_{N}\left\{
\left(E_{0}-E-\frac{\overline{\varepsilon_{N}}}
{W_{N}}\right)^{2}+W_{N}
\frac{\overline{\varepsilon^{2}_{N}}
-\left(\overline{\varepsilon_{N}}\right)^{2}}
{W_{N}^{2}}-\frac{m_{\nu}^{2}}{2}\right\},\\
W_{N}&=&\sum_{n=0}^{N}w_{n},\;\;\;
\overline{\varepsilon_{N}}=\sum_{n=0}^{N}w_{n}
\varepsilon_{n},\;\;\;
\overline{\varepsilon_{N}^{2}}=\sum_{n=0}^{N}
w_{n}\varepsilon^{2}_{n}.\nonumber
\end{eqnarray}
When all the excited atomic states of the ionized molecule are taken
into account, we have $W_{N}=1,\;\;$ and we denote
\beq
\overline{\varepsilon}=\overline{\varepsilon_{N}},
\;\;\;\overline{\varepsilon^{2}}=
\overline{\varepsilon_{N}^{2}},
\;\;\;\sigma^{2}=\overline{\varepsilon^{2}}-
\left(\overline{\varepsilon}\right)^{2}.
\eeq
Now we suppose that there are two components of the source.
According to the calculations \cite{Kap93} %[15]
the value of $\overline{\varepsilon}$ does not change too much with the
composition of the tritiated molecule but the value of $ \sigma^{2}$ can
differ  in the range $\sim$100$\;eV$. So we suppose that
\[\overline{\varepsilon_{1}}=\overline
{\varepsilon_{2}}=\overline{\varepsilon},
\;\;\;\;\sigma_{1}^{2}\neq\sigma_{2}^{2},\]
and we can write for the energies far enough
from the endpoint the following:
\beq
\frac{dP\left(E\right)}{dE}=
\left(E_{0}-E-\overline{\varepsilon}
\right)^{2}+\sigma_{1}^{2}+
B\left(\sigma_{2}^{2}-\sigma_{1}^{2}\right)-
\frac{m_\nu^2}{2}.
\eeq
Here $B$ denotes the amplitude of the second part of the source.
In other words we suppose, for example, that the source composition is
$AT_{2}+B\left(RT\right),\;\;\;
A+B=1$.
The second part with the amplitude $B$, when not taken into account,
can give rise to some counts when compared with  the theoretical spectrum,
calculated for $B=0$. This will be more significant
when being far enough from the endpoint i.e. when all excited states
of ionized molecule are in a play. As concerned the tritium molecules,
these excited states are restricted by $\sim$170$\;eV$. Therefore,
in the integral spectrometer we can expect some rise of the counts if some
-- though small --
admixture (R,T) is present in the source.

\section{Conclusion}
The theme of present work was inspired by
the recently published work of Troitsk group \cite{Lob96} %[9]
on the time dependence of the near endpoint anomaly observed
in the tritium $\beta$-decay. The possibility
that the composition of the tritium source is time dependent due to the
presence  of free T atoms that are
left in the source after the decay of $T_{2}$ molecule was considered.
We suppose that these ions  react with some residual
molecules and remain in the source. It was shown qualitatively that
this can explain some anomalies observed in the electron $\beta$
spectra near the endpoint.

The time dependence of the near endpoint anomaly as reported by Troitsk
group seems to support such a point of view. The presence of this
anomaly as reported in LLNL experiment and the excess of counts in the
lower energy below the endpoint that is observed in all three
experiments can also been qualitatively explained. The differences between
the positions or no observation of the near endpoint anomaly
in Mainz experiment can be also understood. Different tritium sources
used (gaseous in Troitsk and LLNL, condensed film in Mainz) can be the reason.
From our view point the impurities that  can be added little by little
depend on the type of the source due to the different conditions
under which the residual molecules react with tritium ions.

Unfortunately, we are not able to say anything
about the fate of the tritium ions, nor to determine in what a molecular state
supposed impurities can be. Some experimental investigation of
this problem is desirable.
From this viewpoint the recently \cite{Bac96}  %[8]
announced intention to study experimentally the final states
in the tritium $\beta$-decay can be regarded as a first step that can help to
understand this problem.

It seems also desirable to study experimentally another $\beta$-decay
than that of tritium. Unfortunately, the account of  the
final states will be here more difficult than in the case
of the tritium $\beta$-decay.

\vspace{5mm}

{\it This work was supported by Grant Agency
of Czech Republic under the contract No.202/96/0552.}

\end{document}